# A Novel Node Selection Method in Wireless Distributed Edge Storage Based on SDN and Multi-attribute Decision Model

**Yejin Yang[1], Miao Ye[1,*], Qiuxiang Jiang[1], Peng Wen[1]**

[1] School of Information and Communications, Guilin University of Electronic Technology, China

**\* Correspondence:** yemiao@guet.edu.cn;

**Abstract:** The distributed edge storage system can store data collected at the edge of the network in a decentralised manner, with low latency, high security, and flexibility. Traditional edge-distributed storage systems only consider one single factor, such as node capacity, when storing data, ignoring network and storage node load conditions that affecting the system's read/write performance. At the same time, it could be more scalable in the widely used wireless terminal application scenarios. To tackle these challenges, this paper proposes an innovative software-defined edge storage architecture based on SDN (Software-Defined Networking) and SMB (Server Message Block) protocols, A data storage node selection algorithm that integrates the network state and storage node load state is designed based on multi-attribute decision model, and a system prototype is realised in conjunction with 5G wireless communication technology. Experimental results demonstrate significant improvements in the performance of high-load write operations compared to traditional edge-distributed storage systems. The proposed wireless distributed edge storage system also demonstrates superior scalability and adaptability, effectively addressing the challenge of limited system scalability and improving compatibility with edge scenarios in mobile applications. In addition, it results in cost savings in hardware deployment and presents a promising advancement in edge storage technology.

**Keywords:** Edge distributed storage; Software defined networking; Node selection algorithm; 5G wireless communication;

## 1. Introduction

With the continuous advancement of edge computing1, edge storage2, as well as intelligent transportation and Internet of Things (IoT) technologies, an increasing number of intelligent network devices are being deployed at the edge. This surge in deployment has led to a substantial rise in the demand for edge computing and edge storage technologies within edge scenarios, resulting in a dramatic increase in both traffic volume and data requiring processing at the edge 3.

The conventional practice of uploading data to the cloud for processing and storage presents challenges, including risks to user privacy and heightened data transmission delays 4. Addressing these issues, edge distributed storage technology emerges as a compelling solution [5,6]It employs a distributed approach, storing data across multiple edge devices positioned closer to the user. Subsequently, data is selectively transmitted to the cloud center through coding, effectively mitigating concerns related to data security when transmitting sensitive information to the cloud center 7. Simultaneously, this approach alleviates the burden on network bandwidth, enhances data access speed, reduces data transmission latency, and diminishes the risk of a single point of failure.

Additionally, most edge devices are mobile and access the network wirelessly 8. The generated data must be flexibly deployed in the edge storage nodes, requiring that both the edge network devices and the edge storage nodes have wireless communication capabilities for interaction. Edge storage nodes should be easily added, removed, or relocated according to specific needs. This approach provides highly flexible reliability and scalability, ultimately reducing deployment and maintenance costs. The emergence of wireless 5G technology 9 has significantly decreased wireless communication latency and increased data transmission bandwidth. The development of wireless 5G technology has made it possible to combine edge distributed storage with wireless technology, resulting in the edge distributed wireless storage system (EDWS).

In conclusion, researching ways to enhance the overall performance of the edge distributed wireless storage system to cope with the rapid growth of edge traffic holds paramount theoretical and practical significance. It assists in alleviating the bandwidth and storage burdens of the cloud center while safeguarding user privacy data.

In the edge distributed wireless storage system, data is distributed on multiple nodes. Due to the diversity of hardware and software structure of edge storage nodes, edge storage nodes are usually heterogeneous and have different computing and storage capabilities. When data needs to be stored at the edge and a storage node is selected, the computing and storage capabilities of the selected storage node are insufficient, network bandwidth resources are scarce, or the node is faulty or under heavy load. As a result, the data storage and processing efficiency deteriorates. Therefore, node selection is a key issue in edge distributed wireless storage technology 10, which will directly affect the efficiency of data processing and storage and the overall performance of the system.

Presently, node selection in edge distributed storage systems employs several common approaches, including distance-based node selection 11, data type-based node selection 12, and load balancing-based node selection 13. Distance-based node selection involves selecting the nearest node for data processing and storage based on the proximity to the request source. This method reduces data transmission delays, minimizes network bandwidth utilization, and enhances data processing and storage performance. However, it overlooks node computing and storage capabilities, leading to imbalances among storage nodes and affecting system performance.

Data type-based node selection selects the most appropriate node for processing and storing specific data types. For example, video data would be assigned to a node with robust video processing capabilities. This method, however, introduces complexity and computational overhead into the system, as it requires data type analysis. Additionally, it disregards real-time storage node load status and network conditions. When there are significant disparities in data types among nodes, some nodes may become overloaded, resulting in reduced system reliability, performance, and service life.

Node selection based on load balancing aims to identify the optimal node by considering multiple factors, such as load status, computing capacity, storage capacity, and network conditions.

While this method can mitigate the limitations of the previous approaches, traditional edge distributed storage systems require complex and resource-intensive configurations to measure storage node network status [14,15] Moreover, real-time monitoring of storage node load status and computing capacity often necessitates the addition of extra monitoring nodes, increasing deployment costs.

To overcome these limitations and develop an efficient edge storage node selection strategy, it is essential to consider both the state of the edge storage node device and the network status of the edge network. This requires flexible configuration and management of the edge network. Therefore, this paper utilizes the widely adopted SMB protocol 16 as the primary communication protocol and employs a multi-attribute decision method to design and implement an edge-distributed wireless storage node selection algorithm and prototype system based on Software Defined Network (SDN) [17,18].

SDN, as an innovative network architecture, decouples the control plane and data plane in traditional network equipment. This separation allows intelligent and centralized network management through an SDN controller for the control plane and data forwarding via SDN switches for the data plane. The decoupling enhances network control flexibility and programmability, enabling a global perspective on system control. SDN's programmability greatly simplifies network status measurement 19. Furthermore, intelligent use of SDN mechanisms enables real-time monitoring of storage node load status without the need for additional monitoring node hardware. The integration of SDN technology into the edge distributed wireless storage system streamlines the system by reducing hardware configuration and measurement overhead, ultimately enhancing system performance and resource utilization.

The prototype system developed in this study connects to storage nodes via wireless 5G, captures real-time network link status using SDN technology, and continuously monitors storage node load status without requiring additional monitoring node hardware. When clients read and write files, the system considers the current network and storage node load status as weight factors for node selection. It calculates the weight of each storage node using the designed multi-attribute decision model, providing a ranked list of edge storage nodes for data storage.

The main innovations of this paper are as follows:

(1) Compared with the traditional edge distributed storage system with a single data writing method and poor storage capacity scalability, this paper designs an edge distributed storage architecture based on SDN and SMB protocols designed to support wireless 5G communication modes, designs and implements a data storage node selection algorithm in response to the performance and scalability problems of data writing and implements a prototype system of edge distributed wireless storage through physical deployment, which can better cope with the complex edge scenarios, and has the advantages of flexibility in deploying the edge scenarios, low cost of installation and maintenance, and ease of expansion.

(2) Compared with the traditional edge distributed storage system, which only considers the distance between the node and the request source, or the type of the storage file, or the capacity of the storage node as the weight factor of node selection, this paper models the storage node selection problem as a multi-attribute decision-making model, and designs an improved ideal-point solving algorithm. The designed multi-attribute decision model takes into account the real-time network state, real-time load state of storage nodes, and can effectively improve the write performance of the data edge distributed storage system.

(3) Compared with the traditional edge storage system that does not consider the impact of network state on storage performance, this paper realises a new model of software-defined edge storage by combining software-defined network technology in the edge network. By combining the advantages of SDN that can flexibly measure the network state, a storage node load state self-reporting mechanism is designed on the SDN architecture, which can maintain the original system hardware on the basis of the SDN controller's sensing ability, and use the method of constructing packets to obtain the load state of all the storage nodes through the storage nodes' own measurements of the proactive way of reporting, and at the same time, the decision-making algorithms, such as the distribution of data, are uniformly deployed on the SDN controller, without the need to separately add a new model. At the same time, decision-making algorithms such as data distribution are uniformly deployed on the SDN controller, eliminating the need to add a separate dedicated hardware node for collecting the load status of storage nodes, which improves the flexibility of the system configuration and saves the cost of the system at the same time.

The remainder of this paper is structured as follows: Section II provides an overview of previous research on edge-distributed storage systems. Section III describes the data storage process in edge computing, outlines the challenge of the node selection problem, and lists the foundation for the contributions presented in this study. Section IV comprehensively elucidates the architecture of the edge-distributed wireless storage system. It also introduces the proposed storage node selection algorithm and the proposed implementation specifics of the storage node load self-reporting mechanism. The algorithms and implementations of all these designs are under the principle of measuring the state of the network in an SDN-based architecture. In Section V, the paper details the construction of the physical system, which serves as an experimental platform to validate both the effectiveness of the proposed algorithm and the reliability of the overall system through relevant experiments. Finally, Section VI provides a conclusion, highlighting the challenges encountered and presenting potential avenues for future research.[1]

## 2. Related work

This section delineates research pertaining to edge distributed storage systems. The principal distinction between edge distributed storage systems and cloud-centered distributed storage systems lies in their storage locations and usage scenarios. Cloud-centered distributed storage systems are optimal for large-scale data storage and processing, whereas edge distributed storage systems excel in scenarios at the edge where low latency, high reliability, and offline support are imperative.

Rashid 20 et al. designed an edge distributed storage system, EdgeStore, which addresses the problem of inconsistent storage system availability due to the fact that edge devices do not tend to share storage resources and the challenge of configuring all edge devices in a scalable way to integrate with the storage system, by introducing a game-theoretic resource incentives framework, and the whole system is divided into a device allocation module, QoS regulation module, and failure recovery module. However, this edge distributed system does not consider the complex scenarios of mobile edge in its design, using wired deployment is not only costly but also not flexible enough,

---

while only single consideration of storage capacity as the basis of node selection for node gaming in selecting storage nodes is not conducive to load balancing of the system.Makris [21] et al. proposed a lightweight hybrid distributed edge storage framework, by migrating the data close to the farthest user to improve the quality of experience (QoE) of the farthest user, thereby reducing data transfer latency and network utilization. The proposed Edge Storage Component (ESC) utilizes a dynamic lifecycle framework in order to provide containerized applications with transparent and automated access to remote workloads. The effectiveness of the Edge Storage Component is evaluated by using a number of resource utilization and Quality of Service (QoS) metrics. The system is also designed in a wired scenario and still suffers from high deployment costs and lack of flexibility.Sonbol [22] et.al. designed a decentralized distributed storage system, EdgeKV, designed for the edge of the network.Since some distributed storage frameworks and services are usually designed for cloud-centric storage, but the distributed and heterogeneous nature of the edge and its limited resources cause These systems and frameworks are not applicable to the edge, EdgeKV outperforms cloud-centric storage in both local and global data access, and it provides fast and reliable storage by virtue of its location-transparent and interface-based design, utilizing strong consistency features to ensure data replication. This system does not consider how to improve the edge storage performance under the factor of mobile nodes in wireless network scenarios, although it can bring better scalability. Xing [23] et al. proposed a distributed multilevel model for dynamic alternative edge computing storage, in which the storage level consists of the device level on the edge, and when a node's storage space is insufficient, a part of the data is deleted from the current node using the mLFU algorithm, and the date is uploaded to a higher storage level. The system is designed using a multilevel storage model, which requires data migration when some nodes have insufficient storage space, but the wired network approach restricts the flexibility of storage node expansion and raises the cost.Qiao [24] et al. proposed a reinforcement learning based edge distributed trusted storage architecture for solving the problems arising from Intelligent Transportation Systems (ITS, Intelligent Transportation System) in which the heterogeneous transportation network coordination problem arises, uses reinforcement learning based on trustworthiness and popularity to dynamically store data, improve resource scheduling and storage space allocation, and also proposes an authentication protocol based on trapdoor hashing to protect the transportation network access, which effectively solves the problem of data transmission efficiency in ITS. However, the system does not take into account the heterogeneity of storage nodes and the load imbalance brought about when data selects storage nodes for distributed storage.Kontodimas [25] et al. developed a mechanism to perform resource allocation at the edge using a mixed integer linear programming approach and utilized the different characteristics of the edge and cloud resources to store the data in either the edge or the cloud resources, and also utilized corrective censoring techniques to to enhance the system availability and lifetime. The system mainly considers the location of the application and the edge storage node for data placement to increase the access bandwidth and reduce the latency of the edge data, but the method does not consider the real-time load state of the storage nodes and the state information of the network, and the use of only the location information to store the data leads to the problem of load imbalance as well as the problem of over-consumption of the nodes.Wu [26] et al. have designed a new edge de centralized distributed storage system DSPR, by providing cryptographic proofs of stored data, the system can proactively find corrupted data and recover previously unrecoverable data in vulnerable environments.Li [27] et al. address the rapidly growing edge network storage and management problem by proposing an approach that utilizes data popularity (for optimal data access performance) and data similarity (for optimal storage space efficiency) to jointly solve the node selection and data storage problems of edge distributed storage systems. The proposed algorithm is prototyped using Cassandra, an open source distributed storage system, which effectively reduces the service request response time. This method of selecting storage nodes using data types does not comprehensively consider the load of storage nodes as well as the real-time network status, which is not conducive to the load balancing of the system.

The storage node selection strategy and load balancing are very important for the edge distributed storage system, and the result of storage node selection will directly affect the overall performance of the system and device lifetime [28]. The edge distributed wireless storage system designed in this paper uses SDN technology to measure the real-time network state and node load state of the system, which provides more comprehensive decision-making information for these node selection and data placement, reduces the data access latency, and combines with the wireless 5G technology to enable smart devices and storage nodes to interact with the system using a wireless network, which greatly enhances the system's flexibility and scalability. The overall performance of the edge distributed wireless storage system is improved, and the service life of the equipment is extended.

## 3. The designed distributed data edge storage process based on SMB protocol and the storage node selection problem

This section delineates the devised process for distributed data storage at the network edge, leveraging the SMB protocol. It expounds upon the research contributions presented in this paper, encompassing the selection of distributed storage nodes at the edge, state assessment, and system architecture. Notably, given the contextual framework of edge storage, the primary emphasis lies in addressing the storage node selection dilemma. In this context, the wireless storage system designed in this study briefly contemplates the segmentation of a file into multiple smaller entities, subsequently allocating them to distinct storage nodes in a non-redundant manner. It is imperative to underscore that this approach does not impede the applicability of these findings to scenarios involving multi-replica or error-correcting code-based distributed storage configurations.

In conventional edge distributed storage systems, the primary factor considered for selecting storage nodes is their storage capacity. This approach is employed to determine the final priority ranking of storage nodes. However, relying solely on storage capacity as the basis for node selection can give rise to various issues. For instance, disparities in storage capacity among nodes can be significant, and some nodes may experience heavier loads than others. This imbalance has adverse repercussions on system performance, reliability, and service life, resulting in prolonged periods during which certain nodes' storage capacity remains underutilized. This idle state not only wastes storage resources but also diminishes storage efficiency. Furthermore, neglecting to account for the actual network state and the load status of storage nodes can have detrimental consequences. In scenarios of heightened system load, such as during client data read/write operations or system-wide data migration, where the network is congested and storage nodes are under heavy loads, persisting in prioritizing storage node selection based primarily on capacity can severely impact the system's read/write performance and overall storage efficiency.

Henceforth, this study integrates both network state and storage node load factors into the node selection method. The comprehensive process of distributed data edge storage is illustrated in

Figure 1.

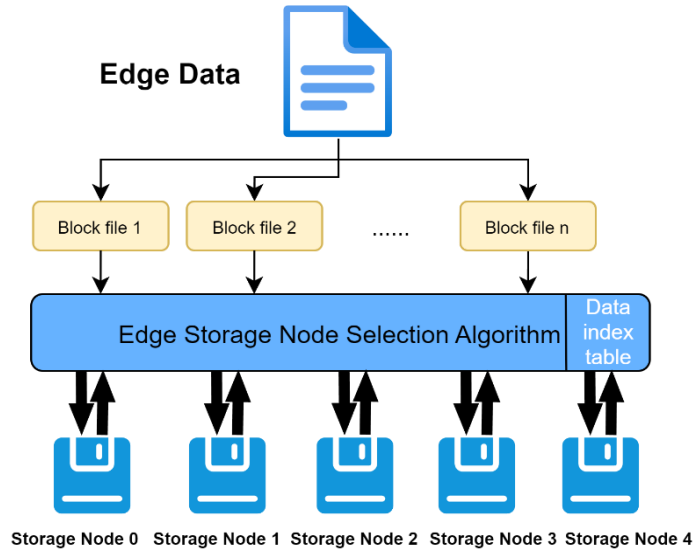

**Figure 1.** Process for data distributed edge storage

Given the premise that edge data is generated to constitute a sizable file, the system initially dissects this substantial file into a series of smaller units. Subsequently, in accordance with the prescribed node selection algorithm, these diminutive files are allocated to a selection of storage nodes possessing sufficient residual storage capacity to meet the stipulated criteria.

Let $M$ be the size of the large file $F$, and the set of suitable storage nodes with available remaining capacity can be expressed as $S = \{s_1, s_2, \cdots s_n\}$, where $n$ denotes the number of storage nodes. The large file necessitates division into $n$ smaller files, where the size of the small file $f_i$, is represented by $m_i$, such that $m_1 + m_2 + \cdots + m_n = M$. A reasonable file division method and the determination of the location of the data block storage obtained after file division need to be associated with the load of the host where the storage node is located and the network load associated with the host, and it cannot be decided only by considering the remaining available storage space of the storage node. For instance, when a storage node is not performing well and the network state is poor, it should be allocated less chunked file storage. Conversely, when a storage node is well loaded and the network is in good state, it should be allocated larger chunks of files for storage. When two storage nodes have similar network state and load state, we should allocate chunks of data of similar size to them for storage. This not only allows the load on the system to be as balanced as possible to reduce the time it takes to write data, but also keeps the spatial distribution of the data as even as possible.

The node selection problem is to determine the size $m_i$ of each small file $f_i$, as well as each small file stored in which storage node $s$, that is, the design of a reasonable node selection mapping mechanism $map$, this mapping mechanism is expressed as the following Equation (1):

$$Node\_list = map(network\_state, node\_state, S) \tag{1}$$

The $node\_list$ represents the list of all storage nodes included in the result of node selection. The mapping function, $map$, is designed to obtain the node selection result by considering multiple factors related to the nodes, which will be explained in subsequent sections. $S$ denotes the collection of storage nodes that satisfy the remaining storage capacity requirement. The $network\_status$ and $node\_status$ are two key factors used in the implementation of the node selection algorithm. The $network\_status$ encompasses three aspects: bandwidth, latency, and packet loss rate, while the $node\_status$ includes four aspects: disk I/O load, remaining storage capacity of the nodes, CPU utilization, and memory utilization. By comprehensively considering real-time network status and real-time storage node load status, the final ranking of storage nodes' superiority and inferiority is obtained. Based on the ratio of superiority and inferiority among all participating storage nodes, the proportion of the input storage file that should be divided is determined, and then it is stored in the corresponding storage nodes. Consequently, the size of the input storage file blocks and the locations for storing the segmented files are determined. With this information, the indexes of the segmented files can be saved, and when file retrieval is required, accessing this index table is sufficient. The consideration of multiple factors can improve the shortcomings of traditional edge distributed storage in node selection, enhance system load balancing, and reduce data response time. The specific storage node selection algorithm will be detailed in Section 4.2 of this paper.

In addition, in the traditional edge distributed storage system architecture, it is very cumbersome to perfor $network\_status$ and $node\_status$ measurements, and it consumes a lot of resources and a huge overhead.The emergence of SDN technology greatly reduces the difficulty and overhead of obtaining these information, therefore, this paper considers the fast and small cost acquisition of $network\_status$ and $node\_status$ information.

Upon completion of the node selection process, the system proceeds to transmit the data blocks of the file to the chosen nodes via the network. For reliable data transmission, conventional network transmission protocols such as TCP/IP [29] are employed. Due to the architectural redesign, some previously utilized data storage communication protocols have become obsolete. This study meticulously examines a range of currently prevalent file transfer protocols, ultimately opting for the SMB protocol [30] as the foundational communication protocol in the system devised herein, with specific implementation using Samba. Once a data block reaches its designated node, it is stored on the node's local storage device, which may include a disk, flash card, or other storage medium. Concurrently, the system maintains a data index to log the file's location and pertinent details. This ensures swift and efficient retrieval of the data block when the file needs to be accessed. This marks the culmination of the entire process of edge distributed storage.

One notable advantage of the edge distributed wireless storage system presented in this study lies in its commendable scalability. Traditional edge distributed storage systems tend to prioritize system reliability while inadvertently neglecting scalability. As the system approaches full capacity or the intricacy of the edge environment intensifies, the original hardware and deployment approach may no longer suffice. At such junctures, the system's adaptability to the evolving edge scenario hinges significantly on its scalability. A primary impediment to scalability often arises from the reliance on wired deployment methods. To address these concerns, this paper introduces a wireless communication protocol within the novel architecture of the edge distributed storage framework. This innovation enables storage nodes and their associated data to interact with the system wirelessly. In instances where expansion of storage nodes is required, they can be seamlessly integrated using wireless 5G technology. This enhancement significantly amplifies the system's scalability, concurrently alleviating the complexities associated with system deployment and maintenance.

## 4. Edge distributed wireless storage system architecture design and algorithm implementation

This section will be divided into four parts. First, it will discuss the overall architecture of the edge distributed wireless storage system designed in this paper. Next, it will introduce the storage node selection algorithm, which is designed based on multi-attribute decision making and incorporates a self-reporting mechanism for storage node load status. Finally, the section will detail the specific implementation of the network state measurement function.

### 4.1 Design of system architecture

The architecture encompasses several essential components, including the control and decision planes, monitoring planes, data forwarding planes, data storage planes, and clients. A comprehensive diagram depicting this structure is presented in Figure 2, sequentially detailing each of these planes.

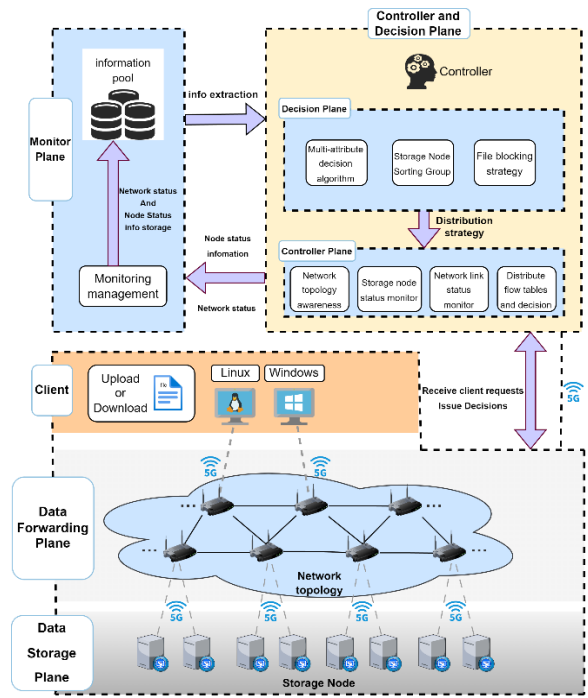

**Figure 2.** Overall frame diagram of edge distributed wireless storage system

### 4.1.1 Control and decision plane

The pivotal component of the system framework is the Controller and Decision Plane (CDP). This element assumes a central role by facilitating network topology discovery 31, retrieving the global network link state and global storage node load state, making determinations on client requests, and overseeing the conduct of the entire network. To achieve comprehensive management, this plane leverages the characteristics of decoupling and centralized control offered by Software-Defined Networking (SDN) in both the control and data planes, thereby utilizing a controller for global oversight.

The controller's role encompasses the measurement and acquisition of comprehensive network link state information. Diverging from prior studies focusing solely on a single network state, this paper comprehensively addresses three pivotal indicators influencing the link state: remaining bandwidth $bw_{remain}$, link latency $delay$, and link packet loss rate $loss$. These factors collectively influence the state of the link from the data source to the storage node.

Merely obtaining network state information falls short of addressing the limitations inherent in the traditional edge distributed storage system. Thus, this paper defines the controller as the central intelligence, responsible not only for gathering network link state information but also encompassing critical load data from storage nodes within the monitoring ambit. In this design, the controller refrains from directly measuring the vital load data of storage nodes; instead, this data is autonomously assessed by the storage nodes and communicated through the load self-reporting mechanism developed in this study. Further details about this mechanism are presented in Section 4.3.

### 4.1.2 Monitor plane

The Monitor Plane comprises a set of functions dedicated to monitoring and overseeing both the performance of the SDN network and the state of node devices within the storage plane. It facilitates real-time monitoring and management of the SDN network's status, performance, traffic, and individual storage plane devices.

Within this paper, the monitoring plane is effectively derived from a segment of the control and decision-making plane. Comprehensive information regarding the network link status and storage node load status is to be regularly and consistently gathered. This data shall find a home in the monitoring plane's information repository, establishing a cohesive and continuous dataset. The monitoring plane is responsible for processing and retaining this information. When decision-making calls for the controller's intervention, this information is retrieved from the repository and supplied to the controller, facilitating appropriate decision-making and the realization of corresponding functions.

### 4.1.3 Data forwarding plane

The Data Forwarding Plane assumes the crucial role of processing and forwarding data, with its central component being the switch. The programmability of SDN fundamentally hinges on the data plane's programmability, exemplified by the OpenFlow Switch [32], a versatile and programmable data plane. This universal programmable data plane empowers users to flexibly define data plane operations via software programming, encompassing packet parsing, processing, and additional functionalities. The data forwarding plane interfaces with both the controller above and the storage nodes below. The controller, situated within the control and decision plane, administers the SDN switch (data forwarding plane) through diverse management modes. These modes encompass in-band management (In-Band) and out-of-band management (Out-of-Band) [33].

The In-Band management (In-Band) mode entails the amalgamation of management data and service data on a singular physical link within the network. This eliminates the necessity for a dedicated management channel, streamlining network management. In contrast, out-of-band management employs a separate physical link for transmitting management data. In In-Band mode, management and service data traverse the same physical link but remain segregated at the software level. This dichotomy ensures separate transmission times, precluding interference. The primary advantage of In-Band mode lies in its substantial reduction of practical deployment costs for edge-distributed wireless storage systems. Notably, this reduction becomes more pronounced with increasingly intricate network topologies. The impact is particularly significant in intricate topologies. If management data traffic is negligible in comparison to business data, its effect on overall network performance can be disregarded. This scenario aligns well with edge deployments.

In conclusion, the controller within the edge-distributed wireless storage system, as designed in this paper, will employ the In-Band approach to oversee the complete data forwarding plane.

### 4.1.4 Data storage plane

The Data Storage Plane bears the responsibility of file storage and retrieval operations. Additionally, it entails the proactive reporting of critical load status information by each individual storage node.

This paper employs the SMB protocol for data storage, implemented using dedicated Samba software. When a client initiates a file storage request and receives the decision outcome from the controller, the file undergoes chunking. Subsequently, the fragmented file interfaces with the relevant storage node through Samba, utilizing multiple concurrent threads to facilitate efficient storage completion.

Due to SDN's inherent configuration, storage nodes are incapable of direct communication with the controller. Thus, the conventional approach to collecting storage node load state involves deploying a host device as a monitoring node. Periodically, this monitoring node establishes connections to storage nodes, retrieves load state information through remote commands, and subsequently stores this data within an information pool. Upon completion, the connection is terminated. When the controller requires a decision based on storage node load state, information is retrieved from the information pool to facilitate decision-making. Notably, this conventional approach is fraught with limitations:

(1) The deployment of a distinct monitoring node substantially amplifies system deployment expenses and enlarges the system's physical footprint.

(2) Frequent interactions between monitoring nodes and storage nodes necessitate continuous controller involvement in sensing and transmitting associated flow tables, consuming a portion of network resources.

(3) In the absence of hardware alterations, storage node load data remains confined to monitoring nodes. Consequently, the controller is unable to access this information. When researchers require this load data for designing decision-making algorithms, their deployment is limited to monitoring nodes. This disperses functionality and diminishes system cohesion.

To tackle the aforementioned problem, this paper ingeniously leverages the packet_in mechanism within the controller [34] to devise an improved solution. This approach circumvents both the issue of network resource consumption by downstream flow tables and the concern of escalating costs without hardware augmentation. Furthermore, it empowers the controller to access information concerning the load state of storage nodes. This integration facilitates the uniform implementation of diverse researcher-designed algorithms within the controller, thereby enhancing system cohesion. The details of this mechanism are expounded upon in Section.

4.1.5 Client

The term "Client" refers to the client software responsible for the practical operation of the distributed wireless storage system at the system's edge. This paper ensures the adaptability of the client software across distinct operating systems by enabling its functional performance on both Linux and Windows platforms. The client software interface, depicted in Figure 3, predominantly encompasses the implementation of two primary functions: file storage and file retrieval. Subsequently, the processes associated with file storage and retrieval are introduced as follows:

File storage:

(1) Users initiate the relevant client software on their Linux or Windows devices, establishing a wireless connection to the SDN switch through a 5G WiFi connection.

(2) Users opt for the storage function, inputting the complete file path (e.g., video file) within the storage window.

(3) The client software automatically computes both the file size and filename. Subsequently, it constructs a packet and appends the file size and filename data to the packet as load data before transmitting it.

(4) Given that none of the switches within the network topology possesses a flow table entry that aligns with the packet, the packet corresponds to the lowest table miss flow table. As a consequence, the controller side initiates a packet_in event, leading to the packet being conveyed to the controller side. Subsequently, an unpacking procedure is undertaken by the controller to extract the file size and file name, thereby culminating in the comprehensive reporting of the information requested by the client for storage purposes.

(5) Upon reception of the storage request from the client, the controller commences the execution of the node selection algorithm (elaborated upon in Section 4.2). Ultimately, the controller attains a result encompassing the file's requisite partitioning into chunks, the individual chunk sizes, and the pertinent IP addresses of the designated storage nodes. Once more, the controller formulates a packet, integrating the obtained result as load data within the packet. This packet is subsequently dispatched back to the client utilizing the packet_out function of the controller.

(6) Upon receiving the decision result from the controller, the client promptly initiates the relevant file chunking program. This program generates a chunking information file in TXT format, designed for future file retrieval, and subsequently activates a multi-threaded concurrent transmission program to dispatch the corresponding chunking file to the designated storage node. At this juncture, the switch lacks a flow table. The controller identifies the connection request between the client and the storage node, promptly accessing comprehensive network link state data within the information pool. Employing the Dijkstra routing algorithm 35, the controller determines an optimally efficient routing and forwarding path, facilitating data transmission from the client (source) to the storage node (destination). This is followed by the execution of the flow table assignment operation. Sequential storage of the individual chunk files culminates in the successful execution of the complete storage function.

File pull:

(1) The client inputs the full path to a segmentation information file that encompasses the filename of the intended file, its total size, as well as the chunked filename, size, and IP address associated with the designated storage node.

(2) Upon clicking the pull button, the client autonomously initiates the reading program, which

endeavors to establish connections with corresponding storage node hosts for sequential file retrieval. During this process, the switch lacks a relevant flow table, prompting the triggering of the controller's packet_in mechanism. Subsequently, the controller promptly accesses comprehensive network link state information stored in the information pool. Employing the Dijkstra routing algorithm based on these link states, the controller ascertains the optimal short route for data forwarding, spanning from the client (source) to the storage node (destination). This culminates in the execution of the operation to dispatch the requisite flow table.

(3) Upon retrieving the chunk file, the file merge program is executed, considering the order of file names (e.g., video_1.mp4, video_2.mp4). The file operation function is utilized to sequentially read each chunk file in binary mode and write it into a new file. Eventually, the complete file name provided within the chunk file's information serves as the designation for the pulled file, ensuring accurate sizing to achieve successful file retrieval.

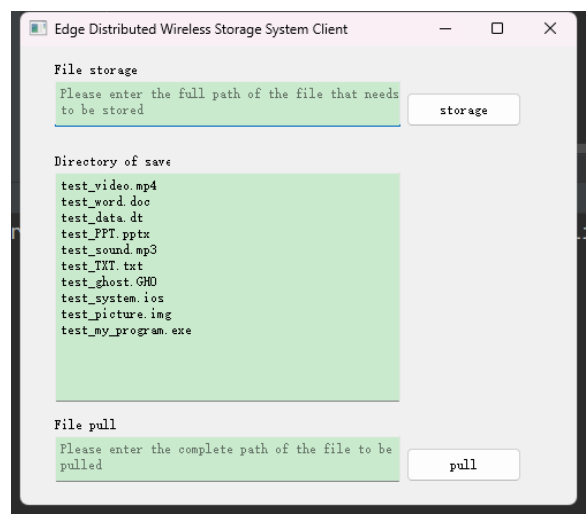

**Figure 3.** Client software interface

*4.2 Edge storage node selection algorithm based on multi-attribute decision making*

The foremost advantage of the edge distributed wireless storage system lies in its capability for distributed file storage. This approach not only ensures high reliability, scalability, and performance, but also provides fault tolerance by dispersing data across multiple nodes. However, conventional edge distributed storage systems base their node selection solely on a single factor, namely storage capacity. This approach, while ensuring spatial distribution uniformity of data, overlooks the crucial influence of both the storage node's internal state and the network condition on storage performance. In scenarios with markedly heterogeneous nodes, some nodes may possess substantial remaining capacity, yet exhibit high load and subpar performance due to their internal state. Despite these shortcomings, their large capacity alone may prioritize their selection, resulting in a pronounced drop in overall storage system performance.

This section introduces the edge storage node selection algorithm grounded in multi-attribute decision-making, developed in this study to rectify shortcomings in the conventional edge distributed storage system's node selection process. The ultimate goal is to enhance the system's write performance.

4.2.1 Selection of decision variables in multi-attribute decision models

A standard edge storage node essentially operates as a computer. Consequently, assessing elements impacting the performance and state of the storage node aligns with evaluating factors affecting a computer's performance and operational load. In order to accurately gauge the performance and load status of a computer, this paper employs the following metrics to represent the influential factors on the node's performance and load state. The specific names and rationales for their selection are delineated as follows:

(1)$L$,disk I/O load, represents the level of input and output operations on the storage node's disk. A high disk I/O load signifies intensive read and write operations. Elevated load can result in delayed response times to client data requests and, in extreme cases, could lead to the failure of the storage node, resulting in data loss. Given that the primary hardware component engaged in client file interaction is the disk, along with other storage devices, the disk I/O load stands as the most significant factor influencing the process of data read and write operations.

(2)$V$,signifies the available disk capacity, indicating the amount of data the node can accommodate. When the remaining capacity approaches full, persisting in file storage may precipitate data loss, severe performance decline, and potential disk damage.This aspect aligns with considerations in conventional distributed storage systems, and is consequently retained as a pivotal factor in this study.

(3)$C$,denoting the CPU utilization rate, serves as a crucial gauge of the available performance capacity of the storage node. Excessive background program operations can lead to a notable surge in the CPU utilization rate, resulting in sluggish data processing. In cases of excessively high utilization rates, system crashes and failures may occur.

(4) $R$,represents memory utilization. While the memory utilization rate does not have a direct impact on client read/write performance, an excess of memory occupation can impede data caching speed on the storage node. Additionally, memory and CPU operations are interdependent. In scenarios involving files with high bandwidth transfer demands, the storage node may necessitate greater memory capacity and lower utilization.

The aforementioned influencing factors were chosen with a comprehensive assessment of the storage node's intrinsic load state. Nevertheless, since data ultimately traverses the network, neglecting to account for network conditions can also impact read/write performance under specific circumstances. For instance, during data migration between storage nodes, there may be an increase in disk I/O load. However, assessing this rise solely based on real-time transmission bandwidth may not adequately reflect the situation. The rise in disk I/O load might be relatively minor, yet the link bandwidth for network transmission could be approaching its limit. Given the rapid advancement of hardware, storage device read and write bandwidth typically surpasses network transmission capacity. Therefore, this study incorporates consideration of both the storage node's load state and network conditions in the selection process. Key metrics for evaluating network state within a network link typically encompass residual bandwidth $bw_{remain}$, $delay$ and packet loss rate $loss$. These three factors have also been incorporated into the input factors for the multi-attribute decision model, ensuring a comprehensive evaluation.

In a complex network topology, numerous potential paths exist from the data source to the storage node, necessitating a decision on which transmission path to employ. To tackle this issue, this paper initially acquires all link states in the network. Subsequently, it applies the Dijkstra routing algorithm, taking parameters like link residual bandwidth $bw_{remain}$, link delay $delay$ and link packet loss rate $loss$ into account, to identify the optimal path from the data source to the target storage node. The data transmission path of this route is considered the node's data transmission path. Extract the remaining link bandwidth $bw_{remain}$, link delay $delay$, and packet loss rate of the path, finally, The values of these three factors are then aggregated to determine the value of the storage node. Finally, the network state of the storage node is computed by summing up the values of these three factors, as specified in Equation (2). Notably, among these factors, the remaining link bandwidth serves as a positive indicator, with a higher value being more desirable. In contrast, the link delay and packet loss rate serve as negative indicators, with smaller values being preferable, and a minimum value of zero.

$$P = scale(bw) - scale(delay) - (loss) \qquad (2)$$

Where scale(x) denotes the function that normalises the variable x.

The selected five influencing factors in this study offer a comprehensive evaluation of both the overall state of storage nodes and the network environment. This is particularly pertinent in cases of node heterogeneity, where it effectively addresses disparities among storage nodes. Consequently, through a thorough assessment of their respective impact on the performance of the edge distributed wireless storage system, the weight ranking of the five indices can be determined, in descending order: disk I/O load, network state P, remaining disk capacity, CPU utilization, and memory utilization. Additionally, a threshold is established for the remaining disk capacity; should it fall below 5%, the node will be excluded from the storage process. Failing to do so may adversely affect read/write performance, potentially resulting in data loss.

### 4.2.2 Multi-Attribute decision node selection algorithm implementation

Upon identifying the five indicators that accurately depict the load of storage nodes and network state, the subsequent step involves a comprehensive evaluation based on the attribute values of each storage node. This evaluation entails sorting nodes based on their state and performance, essentially framing the issue as a multi-attribute decision-making problem within the domain of Mathematical Integration Studies. Multi-attribute decision-making is firmly rooted in systems engineering, with the Technique for Order Preference by Similarity to Ideal Solution (TOPSIS) [36] method emerging as a particularly effective approach. TOPSIS establishes both positive and negative ideal solutions from the normalized raw data matrix, consequently calculating the distances between each solution and these benchmarks. This forms the basis for evaluation.

The primary computational steps of the multi-attribute decision-making algorithm are as follows: initially, the indicators undergo normalization and are assigned weights based on their respective importance, yielding the weighted normalization matrix. Subsequently, the positive and negative ideal schemes are established by identifying the maximum and minimum values of parameter indicators within the weighted normalization matrix. Next, the distances to both the positive and negative ideal schemes are calculated to ascertain relative proximity. Finally, the results are sorted accordingly.

In this study, out of the five chosen factors for assessing storage node status, remaining disk capacity $V$ and network state $P$ are considered positive indicators, with higher values signifying better performance. On the other hand, I/O load, CPU utilization, and memory utilization serve as negative indicators, where lower values indicate improved storage node performance. Within the established multi-attribute decision model, a negative sign is employed to denote negative indicators, as detailed in Algorithm 1. The subsequent discussion will elucidate the algorithm based on the provided diagram.

Step 1 involves operations on Rows 1 to 14 primarily to construct the decision matrix $M$ for storage node weight factors. This matrix is formulated as depicted in Equation (3), where $V$ represents residual capacity, $P$ denotes network state, $L$ signifies disk I/O load, $C$ stands for CPU utilization rate, and $R$ refers to memory utilization rate. Subsequently, the normalized decision matrix $M'$ is derived through the normalization process outlined in Equation (4). Here, $f_{ij}$ represents the elements in matrix $M$, while $i$ and $j$ denote the respective row and column indices, with $n$ representing the total number of storage nodes. The next step involves assessing the number of participating parameter nodes and evaluating the remaining capacity. If the remaining capacity falls below 5%, it holds a veto right, leading to immediate elimination of the storage node. Moreover, in the scenario where only one node engages in storage, further calculations are unnecessary, and the storage node is directly returned.

$$M = \begin{bmatrix} V_1 & P_1 & -L_1 & -C_1 & -R_1 \\ V_2 & P_2 & -L_2 & -C_2 & -R_2 \\ \vdots & \vdots & \vdots & \vdots & \vdots \\ V_n & P_n & -L_n & -C_n & -R_n \end{bmatrix} \tag{3}$$

$$M' = \frac{f_{ij}}{\sqrt{\sum_{i=1}^{n} f_{ij}^2}} \quad i = 1, 2 \cdots n; \ j = 1, 2, 3, 4, 5 \tag{4}$$

In Step 2, spanning Lines 15 to 25, the five factors influencing storage node performance are assigned varying weights, with residual capacity and I/O load bearing considerable significance. To this end, suitable weighting coefficients $w$ are determined, and a normalized weighted decision matrix $Z$ is formulated in accordance with Equation (6). The specific values of the weights in Equation (5) are typically established through a series of experimental iterations.

$$W = \begin{bmatrix} W_V & W_P & W_L & W_C & W_R \end{bmatrix} \tag{5}$$

$$Z = W_j \times M_{ij}' \quad i = 1, 2 \cdots n; \ j = 1, 2, 3, 4, 5 \tag{6}$$

Step 3, encompassing Lines 26 to 30, is dedicated to determining the positive and negative ideal solutions for the weighted decision matrix $Z$, as outlined in Equations (7) and (8).

$$Z^+ = (Z_1^+, Z_2^+, Z_3^+, Z_4^+, Z_5^+) = \max_i \{Z_{ij} \mid j = 1, 2, 3, 4, 5\} \tag{7}$$

$$Z^- = (Z_1^-, Z_2^-, Z_3^-, Z_4^-, Z_5^-) = \min_i \{Z_{ij} \mid j = 1, 2, 3, 4, 5\} \tag{8}$$

Step 4, covering Lines 31 to 34, involves computing the distances $D^+$ and $D^-$ from each storage node to the positive and negative ideal solutions, as demonstrated in Equation (9).

$$D^+ = (D_1^+, D_2^+, \cdots, D_n^+) \qquad D^+ = (D_1^+, D_2^+, \cdots, D_n^+)$$
$$D_i^+ = \sqrt{\sum_{j=1}^{5}(Z_{ij} - Z_j^+)^2} \qquad D_i^- = \sqrt{\sum_{j=1}^{5}(Z_{ij} - Z_j^-)^2} \tag{9}$$

Step 5, spanning Lines 35 to 41, involves the computation of the relative closeness of each storage node to the optimally composed ideal storage node, denoted as $C_i^+$. A higher value signifies superior performance of the respective storage node, as outlined in Equation (10).

$$C_i^+ = \frac{D_i^-}{D_i^+ + D_i^-} \quad i = 1, 2 \cdots n \tag{10}$$

The outlined steps ultimately establish the hierarchical ranking of each storage node within the storage plane.

---

**Algorithm 1 Multiple Attribute Dcision Node Selection**

**Input:**

remaining capacity $V$, disk IO load $L$, CPU usage rate $C$,

memory usage rate $R$, weighting coefficient $W_V$, $W_L$, $W_C$, $W_R$.

**Output:**

relative closeness[ ] - A list of the relative alignment progress between each storage node and the ideal node composed of each optimal decision factor.

**1**   **initialization** $W_V$, $W_L$, $W_C$, $W_R$;

**2**   **initialization** $nodes\_all\_stats\_list$[];

**3**   **initialization** *Checking if the remaining capacity of each storage is less than* 5% *, removing it if true*;

**4**   **initialization** *participate_in_storage_nodes_list*[];

**5**   **if** *The client has sent a request to store a file* $==$ *ture* **then**

**6**      **if** $node\_list.length() == 1$ **then**

**7**        return 1;

**8**      **end**

**9**      **do** *Using* $V, L, C, R$ *construc a weighted factor decision matrix* $M$;

**10**      **for** *each_host_stats* **in** $M$ **do**

**11**        **for** *each_stats* **in** *each_host_stats* **do**

**12**          $M'_{ij} = \sqrt{\sum each\_stats^2}$;

**13**        **end**

**14**      **end**

**15**      ***Obtain*** *normalization matrix* $M'$;

**16**      **for** *each_host_stats* **in** $M'$ **do**

**17**        **for** *each_stats* **in** *each_host_stats* **do**

**18**          $Z_{ij} = M'_{ij} \times W_j$;

**19**          disk_load_list.append(load_value);

**20**          remain_capacity_list.append(capa_value);

```
21                        cpu_utilization_list.append(cpu_value);
22                        mem_utilization_list.append(mem_value);
23                    end
24            end
25            Obtain weighted decision matrix Z;
26            load_best = max(disk_load_list);
27            remain_capa_best = max(disk_load_list);
28            cpu_uti_best = min(disk_load_list);
29            mem_uti_best = min(disk_load_list);
30            for each_host_stats in Z do
```

$$D^+ = \sqrt{\sum_j^{stats\_num}(Z_{ij} - Z_j^+)^2};$$

$$D^+ = \sqrt{\sum_j^{stats\_num}(Z_{ij} - Z_j^+)^2};$$

```
33            end
34            Create all host relative_closeness_list[];
35            for each_host_ in nodes_list do
```

$$C_i^+ = \frac{D_i^-}{D_i^+ + D_i^-};$$

```
37                Relative_closeness_list.append(c_i^+);
38            end
39            return relative_closeness_list;
40    end
```

## 4.3 Design and implementation of a self-reporting mechanism for storage node load status

In this paper, the global network link status is periodically obtained through requests sent by the controller and subsequent measurements. This is an active acquisition method. In contrast, the load status information of all storage nodes is self-measured and reported to the controller by the storage nodes themselves. The communication between the controller and SDN switches is facilitated through the OpenFlow protocol, but direct communication between the controller and the storage nodes connected to the SDN switches is not possible. As mentioned earlier, deploying a separate monitoring node would result in many negative consequences.

To address these issues, this paper presents a mechanism for storage nodes to autonomously report their load status. This mechanism ingeniously utilizes the packet_in functionality within the controller. When a storage node needs to report its load status to the controller, it measures its own real-time disk IO load, remaining disk space, CPU usage, and memory usage through its own commands. It then constructs an empty data packet, encapsulating these load metrics within the packet. Subsequently, this packet is sent with its destination IP pointing to the controller. Since there are no matching flow tables in the SDN switches for this packet, it triggers the packet_in mechanism, which directly sends the packet to the controller. By repeating this process at regular intervals, the controller can periodically obtain load status information from each storage node device, thereby achieving the storage node's autonomous load status reporting mechanism. Additionally, this enables the implementation of a multi-attribute decision-making algorithm on the controller. The operational procedure of the autonomous reporting mechanism is illustrated in Algorithm 2, and we provide an explanation of the mechanism based on the algorithm's flow.

(1) Lines 2 to 7 iterate through all storage nodes, and each storage node autonomously retrieves four key load status indicators via a local Terminal window.

(2) In line 8, using the scapy library 37, an empty TCP packet is constructed, and the values of the four key status indicators are appended to the packet as payload.

(3) Lines 9 to 10 employ the storage node's IP as the source IP for the packet and the controller's IP as the destination IP, followed by packet transmission.

(4) Lines 11 to 12 indicate that the packet cannot be matched with any pre-existing flow tables in the switches, thus immediately triggering the packet_in event, which forwards the packet to the controller. The controller, at this point, does not initiate flow table actions but rather stores the packet.

(5) Lines 13 to 17 describe the controller-side process of parsing the packet, extracting the payload data containing the values of the four key status indicators. After a 3-second delay, it proceeds to the next data acquisition cycle, continuing in an infinite loop, completing the storage node's autonomous reporting process.

Should the controller need to make decisions and issue decision results, it can use a reverse approach. An empty TCP packet is constructed, with the decision result appended as payload. The controller then uses the packet_out functionality to specify the SDN switch port connected to the client requesting file operations, facilitating the forwarding of the decision result to the client.

---

**Algorithm 2** Storage Node Self-Reporting Mechanism

---

**Input**:All storage node objects

**Output**: Send all storage node load status information to the controller

**1  While** *True* **do**:

**2**    **for** *each_nodes* **in** *all_nodes_list* **do**

**3**       Storage nodes: get remaining capacity $V$ , disk IO load $L$, CPU usage rate $C$, memory usage rate $R$;

**4**       Storage nodes: Terminal(df -lm, obtain $V$) ;

**5**       Storage nodes: Terminal(iostat -x 1 -t 3, obtain $L$);

**6**       Storage nodes: Terminal(top -bn 1 -i -c, obtain $C$);

**7**       Storage nodes: Terminal(cat /proc/meminfo, obtain $R$);

**8**       Storage nodes: Construct an empty TCP packet, loading $V$, $L$, $C$, $R$;

**9**       Storage nodes: set packet.src_ip = nodes_ip, packet.dst ip = Controller_ip;

**10**      Storage nodes: send(TCP packet);

**11**      Controller: trigger Controller.ofp_event.EventOFPPacketIn;

**12**      Controller: save TCP packet;

**13**      Controller: analysis TCP packet, Unloading load data;

**14**      Controller: obtain this nodes $V$, $L$, $C$, $R$;

**15**    **end**

**16**

**17**    **delay 3s**;

*4.4 Implementation of node load and network status measurement functionality*

The storage node operating system designed in this study is Linux. The measurement of four states of node load is relatively straightforward. Among them, the disk IO load can be obtained by continuously reading with the 'iostat' tool for 3 seconds. Remaining disk capacity can be acquired in Linux using the 'df -lm' command. Similarly, CPU usage and memory usage can be obtained by using 'top -bn 1 -i -c' and 'cat /proc/meminfo' commands, followed by the application of regular expressions.

In the measurement of residual bandwidth in link, it is assumed that the current need is to measure the link bandwidth between switch $swA$ and switch $swB$. Let the current moment be denoted as $t-1$, The controller simultaneously issues a request to both $swA$ and $swB$. Upon receiving the Request, $swA$ and $swB$ immediately return their respective total sent byte count $tx$, total received byte count $rx$, and the timestamp $time_{t-1}$ corresponding to the current time $t-1$ to the controller. After waiting for approximately 1 second, denoted as time $t$, the controller issues another request, allowing the retrieval of $tx$, $rx$, and $time_t$.For calculating the utilized bandwidth of $swA$ during this time interval from $time_{t-1}$ to $t$, the total bytes transmitted and received at time $tx+rx$ are subtracted from the corresponding values at $time_{t-1}$, yielding the total traffic for $swA$ during this period. Then, according to Formula (11), the utilized bandwidth of $swA$ can be determined. As indicated in Formula (12), $swA$ remaining bandwidth is calculated by subtracting the utilized bandwidth from the maximum link bandwidth of the switch port, resulting in $swA$ remaining bandwidth. Similarly, the remaining bandwidth for $swB$ can be determined.In this study, a rigorous approach is employed for all data calculations, whereby the link bandwidth between $swA$ and $swB$ is determined as the minimum value between the two, as per the formula (12).

$$bw_{used} = \frac{(tx_t + rx_t) - (tx_{t-1} + rx_{t-1})}{time_t - time_{t-1}} \tag{11}$$

$$bw_{remain} = bw_{max} - bw_{used} \tag{12}$$

In the measurement of link latency, the controller issues a Packet_out message to $swA$. The data segment of the message contains the timestamp at the time the controller issued the message. The action specified in the Packet_out message instructs the switch to either flood it or forward it to a specific port. Until $swB$ receives the data packet sent by $swA$, and no corresponding flow table entry can be matched, triggering a Packet_in to the controller. Upon receiving this data packet, the controller calculates the time difference $T_1$ by subtracting the current time from the timestamp in the received packet. This time difference is approximately equal to the latency from the controller to $swA$, plus from $swA$ to $swB$, and finally from $swB$ back to the controller, as illustrated in Figure 4.Similarly, the controller performs the reverse operation to obtain the time difference $T_2$, as shown in Figure 4. The controller then sends Echo request messages to $swA$ and $swB$, each containing the current timestamp. Upon receiving these messages, the switches immediately reply with reply packets carrying the same timestamp. By subtracting the timestamp in the reply packet from the time of packet reception, the controller can determine the round-trip latency $RT_A$ and $RT_b$ between the controller and $swA$ as well as between the controller and $swB$, as depicted in Figure 5. Assuming the round-trip latencies are equal, Formula (13) can be used to calculate the link latency $Delay_{AB}$ between $swA$ and $swB$.

$$delay_{AB} = \frac{T_1 + T_2 - RT_A - RT_B}{2} \tag{13}$$

The measurement method for link packet loss rate is similar to that of link residual bandwidth measurement. Taking $swA$ as an example, the controller periodically issues requests to obtain the total number of bytes sent $swA_{tx}$ and received $swA_{rx}$ by $swA$ during the time interval $T_2 - T_1$. Similarly, the total sent bytes $swB_{tx}$ and total received bytes $swB_{rx}$ for $swB$ can be determined. Let's denote $loss_{AB}$ as the packet loss rate in the direction from $swA$ to $swB$. According to Formula (14), the final link packet loss rate between $swA$ and $swB$ can be calculated.

$$loss_{AB} = max(1 - \frac{swB_{rx}}{swA_{tx}}, 1 - \frac{swA_{rx}}{swB_{tx}}) * 100\% \tag{14}$$

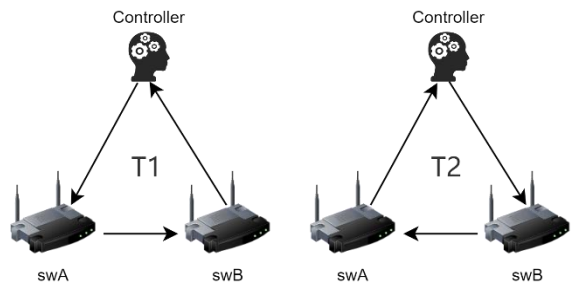

**Figure 4.** The controller sends Packet_out packets

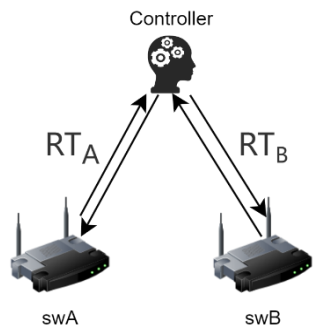

**Figure 5.** The controller sends Echo request packets

## 5. Implementation and Testing of an Edge Distributed Wireless Storage System

### 5.1 Experimental platform design and realization

To validate the efficacy of the suggested multi-attribute decision-based node selection algorithm, integrated with the node load self-reporting mechanism within the context of the edge distributed wireless storage system, this section assesses the influence of the devised system and algorithm on the system's read/write performance using a real-world experimental platform.

In contrast to conventional edge distributed storage systems, this study strategically selects and deploys hardware to suit the edge-end scenario, thereby distinctively differentiating all hardware components from other systems. This alignment with the edge-end scenario is emphasized. The prototype system encompasses 1 controller, 3 storage nodes, and 14 switches. The software ecosystem encompasses: OpenWrt 38 version 22.03, Open vSwitch 39 version 2.1.3, Ryu version 4.8, OpenFlow version 1.3, and Samba version 4.13. The controller operates on Ubuntu 20.04 LTS, while the storage nodes operate on Debian 10. The subsequent section offers detailed insights into the hardware design and implementation, accompanied by an analysis of the conclusive experimental outcomes.

### 5.1.1 Controller hardware design and implementation

Within the realm of SDN, controllers employ two primary control methods for switches: single-controller control and multi-controller control. While this paper primarily delves into single-controller control, it acknowledges the impending system upgrade to multi-controller control, where the enhancement algorithm amplifies system performance. Consequently, emphasis is bestowed upon multi-port hardware devices. Moreover, given the edge positioning of the system, compact device dimensions are paramount, while performance remains a prerequisite.

Based on the aforementioned criteria, this study ultimately opts for the N5105 six-network port soft routing host from the Smooth Network Micro-Controller Company to serve as the hardware platform for the controller. The soft routing host boasts six 2.5G wired ports and wireless WiFi functionality. Its compact dimensions, measuring a mere 17.8cm (length) * 12.5cm (width) * 5cm (height), align seamlessly with the edge-end system design paradigm. The physical depiction of the host is illustrated in Figure 6.

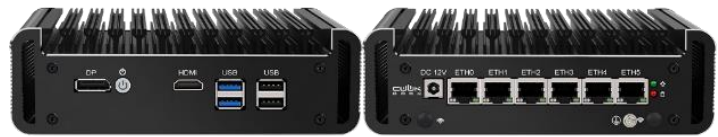

**Figure 6.** Physical drawing of N5105 soft route mainfram

### 5.1.2 SDN Switch hardware design and implementation

The role of an SDN switch is to manage data forwarding within the physical system, and its performance attributes significantly impact the overall network performance. Presently, SDN switches available in the market cater primarily to expansive cloud storage centers due to their size, rendering them less suitable for mobile applications. Furthermore, the cost of these SDN switches is substantial, often reaching tens of thousands of dollars per unit. As a result, this study confines the hardware selection for SDN switches to commonplace home routers. These routers are compact in design and generally cost-effective, aligning effectively with real-world edge scenarios. Nonetheless, it's important to note that while home routers are used, they do not inherently possess the capabilities of SDN switches. To harness SDN functionality, a transformation process is requisite.

OpenWrt, a Linux-based open-source embedded operating system [38], is purpose-built for routers and embedded devices. It offers a versatile, customizable, and expandable platform, capable of replacing stock firmware to enhance functionality and control for routers and embedded devices. Open vSwitch (OVS), an open-source virtual switch software [39], serves as a tool to construct and oversee networks within virtualized environments. It furnishes an extensive array of network features and management utilities for establishing connections with physical networks. The combination of OpenWrt and Open vSwitch constitutes pivotal components in the process of converting routers into SDN switches.

To convert a router using the aforementioned components, several factors must be considered. Firstly, the router should fall within the spectrum of models supported by the OpenWrt system. Secondly, in real-world edge distributed storage scenarios, SDN switches grapple with substantial real data traffic and data forwarding demands. Following the router's transformation, the original dedicated data forwarding chip might encounter performance limitations, or in some cases, become unusable. Lastly, upon integration with Open vSwitch, the router's background resource consumption experiences a significant surge. Should the chosen router possess inadequate performance capabilities, potential consequences encompass tardy data forwarding responses, excessive resource usage, heightened heat generation, and even system crashes.

To address potential issues arising from router modifications, this study has selected the XiaoMi AX6000 router as the physical SDN switch for this system, as depicted in Figure 7. This router boasts a robust 2GHz CPU frequency, generous RAM and ROM capacities, as well as a compact form factor, aligning seamlessly with the system's edge-end positioning and design philosophy. The critical performance metrics of the router are provided in Table 1. The transformation of the router into an SDN switch is achieved via successive stages, including OpenWrt source code modification, firmware compilation, overriding the original system with burn-in procedures, and installing the Open vSwitch plug-in. This process adeptly equips the router to fulfill SDN switch functions.

**Table 1.** XiaoMi AX6000 Router Performance Parameters

| model number | CPU Model | CPU frequency | Flash size |
|---|---|---|---|
| AX6000 | MT7986A | 2GHz | 128MB |

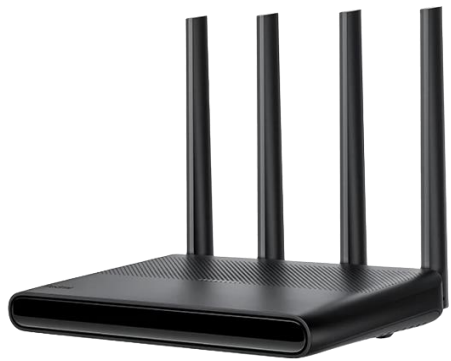

**Figure 7.** XiaoMi AX6000 router

### 5.1.3 Edge storage node hardware design and implementation

Hardware design for storage nodes should align with authentic edge-side scenarios. In practical settings, storage nodes often exhibit heterogeneity, a key distinction from simulation systems, attributed to the following factors:

1) Within identical equipment types, inherent process variations result in certain performance disparities; complete uniformity of performance cannot be attained.

(2) Divergent installation sites for storage node devices, varying environmental conditions, and external factors such as heat dissipation lead to progressive performance divergence among these devices over time.

(3) In the context of a deployed edge distributed storage system, which may undergo equipment updates and iterations, new and existing equipment coexist within the same distributed storage system, giving rise to a situation of storage node heterogeneity.

Based on the preceding analysis, the heterogeneous nature of storage nodes aligns better with real-world edge scenarios. Consequently, this paper adopts a heterogeneous design for the storage nodes. When selecting hardware, it becomes imperative to opt for devices featuring varying levels of performance to achieve this heterogeneity.

After a thorough comparison, three devices—Asus Tinker Board 2S (No.1), Raspberry Pi 4B (No.2), and Raspberry Pi 3B+ 40 (No.3)—have been chosen as the designated hardware platforms for the storage nodes in the edge-distributed wireless storage system proposed in this paper. The devices are ranked in order of performance: No.1, No.2, and No.3. Their essential parameters are tabulated in Table 2, while their physical embodiments are depicted in Figure 8.

**Table 2.** Comparison of performance parameters of three embedded single board computers

| Equipment Model | CPU Model | CPU frequency | RAM type | RAM size |
|---|---|---|---|---|
| Tinker Board 2S | RK3399 | 2GHz | DDR4 | 4GB |
| Raspberry 4B | BCM2711 | 1.5GHz | DDR4 | 2-4GB |
| Raspberry 3B+ | BCM2837 | 1.4GHz | DDR2 | 1GB |

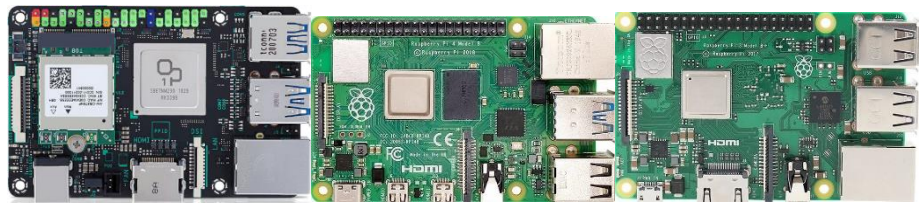

**Figure 8.** Physical diagrams of the three storage nodes

### 5.1.4 Network topology design and physical deployment

To enhance complexity and replicate a highly realistic edge-end setting, this study configures 14 switch nodes, thereby establishing the comprehensive network framework illustrated in Figure 9. Interconnecting the devices, all SDN switches establish wired connections with each other to bolster reliability. Moreover, while the physical SDN switches are equipped with hardware provisions for wireless networking capabilities, direct utilization is unfeasible. Additionally, an examination of the topology reveals the presence of multiple loop-formed network links, which could potentially trigger broadcast storms [41].

This paper addresses the aforementioned issues by implementing Virtual Local Area Network (VLAN) isolation technology [42]. VLAN divides the physically established LAN into distinct logical subnets, isolating data link layer broadcast messages within these subnets to create individual broadcast domains. Each logical subnet constitutes a VLAN. Terminal devices accessing the VLAN are assigned to specific VLANs, preventing direct communication between devices in different VLANs through the data link layer. By utilizing VLAN technology, the scope of broadcast message transmission is confined, mitigating the impact of broadcast storms and enhancing network security. Employing VLAN, the wireless network card within the physical SDN switch can be partitioned into separate VLANs and integrated into the SDN switch, enabling devices beyond the SDN switch to join the edge distributed wireless storage system via wireless network connections. Each connected device operates within its own distinct LAN, preventing mutual interference and effectively mitigating broadcast storm issues within the network.

In the implemented system outlined in this paper, the controller devices, storage nodes, and client devices establish connections to the SDN switch through WiFi wireless 5G connections. This design choice brings the system in closer alignment with edge-end environments, enhancing system scalability. The tangible manifestation of the implemented configuration is depicted in Figure 10.

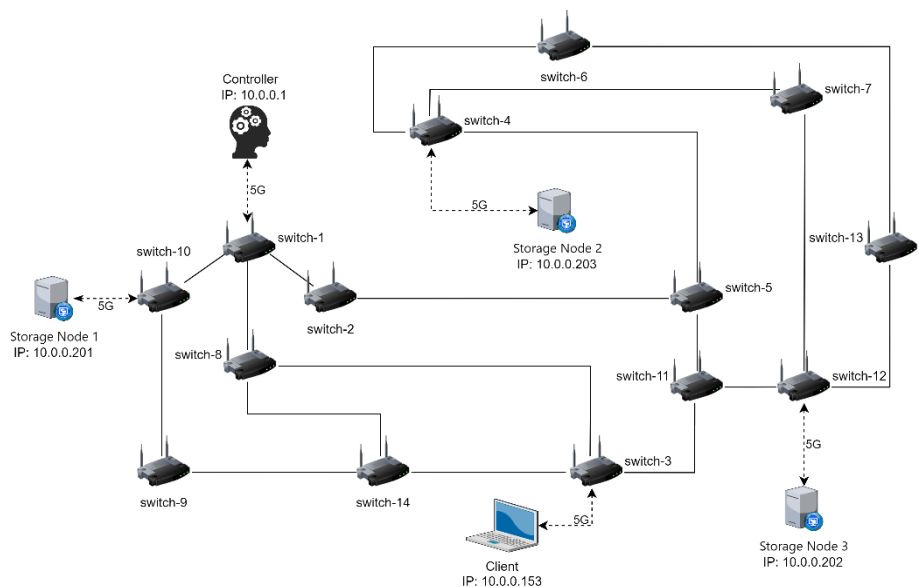

**Figure 9.** Network topology

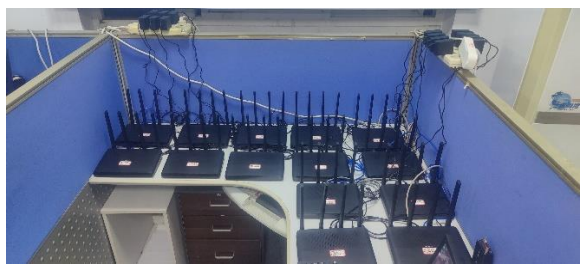

**Figure 10.** Overall physical picture

*5.2 Performance testing and analysis of results*

In order to verify the performance of the system for reading and writing objects of different sizes, three sizes of files, namely 10MB, 100MB, and 1000MB, are set for storage tests, and 30 storage tests are conducted for the same size of files in each round. Client from the moment of clicking the request storage as the start time, the file storage time as the end time, the two can be obtained by subtracting the total time consumed to store the file, through the design of this paper's edge-distributed wireless storage system (EDWS) and the traditional edge-distributed storage system (TEDS) compared to get the results of the test.

The overall experimental phase is divided into two parts. The first part is the storage time comparison test for different file sizes when the network is not congested and the nodes are in normal state; the second part is the storage time comparison test for different file sizes when the network bandwidth resources are reduced and some of the nodes are in high load state. The second part is the storage time comparison test for different file sizes when the network bandwidth resources are reduced and some nodes are under high load. This testing method can comprehensively reflect the performance of the system in the normal state and stress state.

The first is the test with normal network state and node load, where the horizontal coordinate of Figure 11 is the number of tests and the vertical coordinate is the total storage consumption time. Figure 11(a) is a comparison chart of the storage time tested in two different systems under the condition of file size of 10MB respectively, where the green line is the test result of the system designed in this paper, and the red line is the test result of the traditional edge-distributed storage (without SDN), and it can be seen from the experimental results that, in terms of the storage of small files, although there is not a big difference in the time effect of the two comparisons, the overall view of the system designed in this paper storage time is lower than the traditional edge distributed storage system. Figure 11(b) is a comparison chart of the storage time tested in two different systems under the condition of file size of 100MB respectively, from which it can be seen that the storage time of the system designed in this paper is lower than that of the traditional edge-distributed storage system for the first 15 times, and from the 15th time onwards, there will be occasional increase in time, which is due to the fact that the transformed SDN switch does not mobilize the dedicated data forwarding chip aggressively , the long running time leads to heat generation, which makes the switch processor downclocked and slows down the processing data resulting in longer time. Figure 11(c) is a comparison chart of the storage time tested in two different systems under the condition that the file size is 1000MB respectively, comparing the three charts to see that with the increase of the storage file size, the storage time folds of the two systems are gradually separated, which means that the larger the file to be stored, the advantage of the edge-distributed wireless storage system designed in this paper will be more obvious, this is because although the system is in the network state is good, but due to the long operation of some physical switches to generate heat, frequency reduction and other reasons, will lead to some switch nodes appear transient processing slow phenomenon, this time the controller on all the links of the bandwidth, latency, packet loss rate measurements, will find this phenomenon, so that according to these measurements out of the data re-decision-making, issued a new flow table to let the data through the better link for transmission, thus reducing the performance degradation caused by hardware problems.

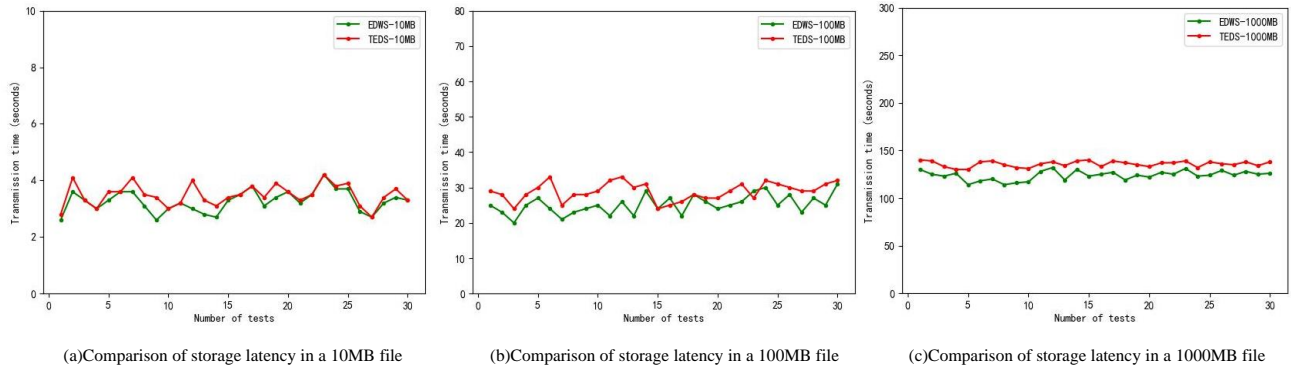

(a)Comparison of storage latency in a 10MB file      (b)Comparison of storage latency in a 100MB file      (c)Comparison of storage latency in a 1000MB file

**Figure 11.** Comparison of storage latency in a 10MB,100MB,1000MB file

The second part of the experiment, is a test in the case of poor network status and increased load on certain nodes, here a program will be written to implement a looped video playback in storage node No.2 (Raspberry Pi 4B) to increase the CPU and memory usage of this storage node. And during the test, separate file storage is performed in storage node No.2 through other clients to increase the disk IO load of this node and reduce the state of the network. Figures 12 demonstrate the experimental result graphs for this condition.

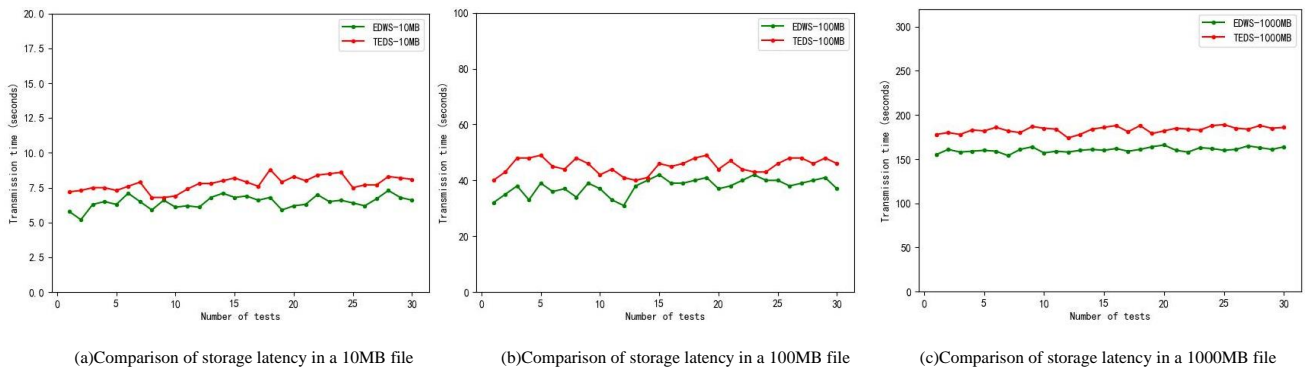

(a)Comparison of storage latency in a 10MB file      (b)Comparison of storage latency in a 100MB file      (c)Comparison of storage latency in a 1000MB file

**Figure 12.** Comparison of storage latency in a 10MB,100MB,1000MB file

As can be seen from the figure, when the network state is poor and the load of storage node No. 2 is large, the storage time of the three different file sizes increases to a certain extent, but the system designed in this paper still has a better performance and a shorter storage time than the traditional edge-distributed storage system, and the advantage is more obvious than that of the case where the network and the storage node are well loaded. This is due to the fact that the edge-distributed wireless storage system designed in this paper does not only take the remaining capacity of the storage node as an index when selecting the storage node, but also considers the state of the network and the load state of the storage node comprehensively, when the performance of the No.2 storage node and the state of the network decreases, the controller obtains the network state and the load state of the No.2 storage node before doing the file chunking and the selection of the storage node, and comprehensively judges for this storage node state is poor, then the decision of splitting files will allocate more file size to storage nodes 1 and 3, reducing the file storage size of storage node No.2 and increasing the overall system performance. The performance of the edge distributed wireless storage system designed in this paper is verified by the above two parts of experiments.

## 6. Conclusion.

In this paper, an edge distributed wireless storage system architecture based on SMB and software-defined networking is designed and a prototype physical system is fabricated. The system utilizes software-defined networking technology to sense and measure the real-time network link state, and at the same time obtains the real-time load state of the nodes through the storage node load state self-reporting mechanism, combines the two information to establish a multi-attribute decision model, and completes the selection of storage nodes by solving the model. Through the write performance test of the actual system, compared with the traditional edge distributed storage system, the system designed in this paper has obvious improvement in write operation performance; and compared with the edge distributed storage system that needs to deploy monitoring nodes individually, it has the advantages of light weight and low deployment cost. It is believed that the work in this paper can help promote the development of edge distributed storage systems.

In addition, in real edge scenarios, a large number of sensors and collectors are generating data, which involves the complexity of data, storage and change of duplicated data, and the network topology will be more complex, which will lead to the problems of data consistency, accuracy and validity [43]. In addition, the routing decision of data is also a hot topic in current research, and choosing a better data storage path in the complex edge network is very helpful to improve the overall performance of the system. How to adopt better consistency protocols and routing algorithms to solve the above problems is the direction that will be studied in the next work of this paper.

## Acknowledgments

This work was supported in part by National Natural Science Foundation of China (Nos.62161006, 62172095), the subsidization of Innovation Project of Guangxi Graduate Education (No. YCBZ2023134), Key Laboratory of Cognitive Radio and Information Processing, Ministry of Education (Guilin University of Electronic Technology) (No. CRKL220103), and Guangxi Key Laboratory of Wireless Wideband Communication and Signal Processing (Nos. GXKL06220110, GXKL06230102).

## Conflict of interest

The authors declare there is no conflict of interest.